\RequirePackage[2020-02-02]{latexrelease}
\documentclass[aps,prb,twocolumn]{revtex4}
\usepackage{ucs}
\usepackage{graphicx}
\input epsf
\usepackage{amsmath}
\usepackage{amssymb}
\usepackage{amsfonts}
\usepackage{mathrsfs}
\usepackage{float}
\usepackage{color,soul}

\begin{document}
\title{Gate-tunable crossover between vortex-interaction and pinning dominated regimes in Josephson-coupled Lead-islands on graphene}
\author{Suraina Gupta}
\affiliation{Department of Physics, Indian Institute of Technology Kanpur, Kanpur 208016, India}
\author{Santu Prasad Jana}
\affiliation{Department of Physics, Indian Institute of Technology Kanpur, Kanpur 208016, India}
\author{Rukshana Pervin}
\affiliation{Department of Physics, Indian Institute of Technology Kanpur, Kanpur 208016, India}
\author{Anjan K. Gupta}
\affiliation{Department of Physics, Indian Institute of Technology Kanpur, Kanpur 208016, India}

\date{\today}

\begin{abstract}
Resistance of a Josephson junction array consisting of randomly distributed lead (Pb) islands on exfoliated single layer graphene shows a broad superconducting transition to zero with an onset temperature close to the transition temperature of bulk Pb. The transition evolves with the back-gate voltage and exhibits two peaks in temperature derivative of resistance. The region above the lower temperature peak is found to be well described by Berezinskii-Kosterlitz-Thouless model of thermal unbinding of vortex anti-vortex pairs while that below this peak fits well with the Ambegaokar-Halperin model of thermally-activated phase slip or vortex motion in Josephson junction arrays. Thus a gate-tunable crossover between interaction and pinning dominated vortices is inferred as the Josephson energy, dictating the pinning potential magnitude, increases with cooling while the effective screening length, dictating the range of inter-vortex interaction, reduces.
\end{abstract}

\maketitle

\section{Introduction}
Superconductivity (SC) in two dimensions (2D) has been intriguing for many reasons including the abundance of low energy fluctuations due to the absence of long range order in 2D at finite temperatures \citep{kadin1983renormalization,newrock2000two}, presence of superconductor to insulator transition (SIT) \citep{allain2012electrical,hen2021superconductor,li2010transport,bollinger2011superconductor} and a super-insulator phase \cite{baturina2013superinsulator}. The granularity plays an important role in 2D-SC and the interplay between Coulomb energy $E_{\rm C}$, Josephson coupling energy $E_{\rm J}$ and the thermal energy $k_{\rm B}T$ broadly dictates different regimes. The Coulomb energy $E_{\rm C}$ is the energy cost of exchanging a Cooper pair with a SC island \cite{van1996quantum} while $E_{\rm J}$ dictates the energy cost for two SC grains to have unequal phases. The $E_{\rm C}$'s dominance leads to a super-insulator phase with localized Cooper-pairs and mobile vortices. A competition between $E_{\rm C}$ and $E_{\rm J}$, with both dominating over $k_{\rm B}T$, leads to SIT while the dominance of $E_{\rm J}$ leads to SC.

For $E_{\rm J} \gg E_{\rm C}$, the low energy phase-fluctuations, manifest as vortex anti-vortex pairs, at finite temperatures. This broadens the SC transition in a 2D-SC and it has mostly been described by the much celebrated works, of Berezinskii \cite{berezinskii1971destruction} and, Kosterlitz and Thouless \cite{kosterlitz2018ordering}, known as the BKT model. In this model the thermal unbinding of such pairs lead to a finite resistance, at any non-zero current, below the bulk SC critical temperature $T_{\rm C}$ and above a phase-transition temperature called $T_{\rm BKT}$.

As compared to a 3D-SC, the screening capability of currents in a 2D-SC for perpendicular magnetic fields is much weaker \cite{pearl1964current} and so the effective screening length $\lambda_{\perp}$, which dictates the inter-vortex interaction, can even exceed the sample size. Eventually, the logarithmic dependence of the interaction energy on the inter-vortex separation, required for BKT model, is possible only for sample sizes below $\lambda_{\perp}$. This imposed finite sample-size rules out a true BKT phase-transition in thermodynamically large 2D samples. Nevertheless, in certain regimes signatures of the BKT transition have been observed \cite{halperin1979resistive,sun2018double,schneider2009electrostatically}. With cooling, $\lambda_{\perp}$ can become smaller than the sample size and, in some cases, this may happen at temperatures above the anticipated $T_{\rm BKT}$.

Another regime of interest in a finite size 2D array of JJs is described by Ambegaokar-Halperin (AH) model of thermally activated phase slips (TAPS) \cite{ambegaokar1969voltage,tinkham1988resistive}. In this case, the vortices independently drift, under a bias current, through the inter-grain regions and over a potential landscape \cite{rzchowski1990vortex}, dictated by $E_{\rm J}$ and grain distribution, leading to finite resistance. The vortex anti-vortex pairs with small separation nucleate in the bulk as thermal excitations which then exit from the opposite edges after drifting under the influence of the bias current.

Graphene's exposed two-dimensional electron gas with an easy control of its carrier density through back-gate makes it a popular tunable substrate for studying gate tunable SC. The SC materials that do not wet the graphene surface are ideal for this as one can easily get a 2D array of JJs by using suitable deposition conditions. Many interesting phenomena including SIT \citep{allain2012electrical,sun2018double} have been observed in such SC-graphene hybrid systems.

In this paper, we report on resistance vs temperature $R(T)$ and current voltage characteristics (IVCs) of a four probe device consisting of lead (Pb) islands on exfoliated single layer graphene with back-gate voltage providing a handle on $E_{\rm J}$. The resistance shows a broad gate-dependent transition to zero resistance with two peaks in its temperature derivative. The gate voltage dependent $R(T)$ is well described by the BKT and AH models for temperatures above and below the low temperature peak, respectively. The IVCs in the low temperature region are also found to be consistent with the AH model. Finally a gate-tunable crossover between interaction and pinning dominated vortex regimes is concluded.

\section{Experimental Details}
\begin{figure}[h!]
	\centering
	\includegraphics[width=3.4in]{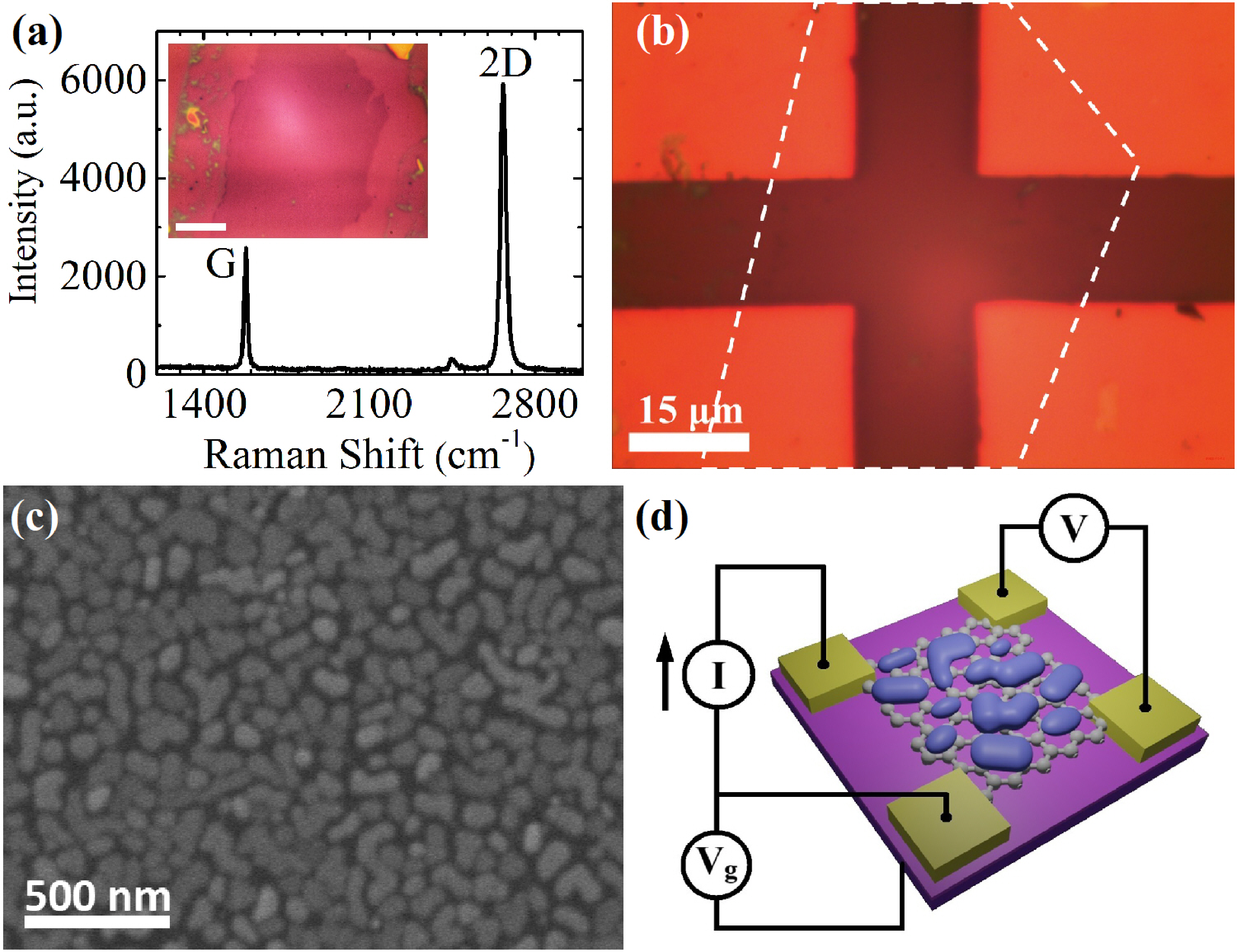}
	\caption{(a) Raman spectrum taken on graphene shows the two characteristic Raman peaks at approximately 1577 ${\mathrm{cm^{-1}}}$ (G peak) and 2665 ${\mathrm{cm^{-1}}}$ (2D peak). The ratio I(2D)/I(G) = 2.3 confirms the graphene to be single layer. The inset is the optical image of single-layer graphene (SLG). (b) Optical image shows the four contact pads in a vander pauw geometry. The white dashed line marks the graphene boundary. (c) An SEM image showing the distribution of Pb islands (grey color) on graphene (black background). (d) Schematic diagram of the four terminal configuration for electrical measurements in the Pb-graphene hybrid device.}
	\label{fig:GrPb1}
\end{figure}

Monolayer graphene was prepared by exfoliating Kish graphite on highly p-doped Silicon wafers with 300 nm thick gate-quality oxide on it. The substrate was first cleaned by sonicating it in acetone, isopropyl alcohol (IPA) and de-ionized water, consecutively for 5 minutes in each, and then in 50 W oxygen plasma for 2 minutes \cite{huang2015reliable}. Exfoliation was done within 30 minutes of this cleaning process. Graphene flakes were then identified under optical microscope. The inset in Fig. \ref{fig:GrPb1}(a) shows the graphene monolayer used for fabricating the actual device. Raman spectrum of the graphene flake, in Fig. \ref{fig:GrPb1}(a), gives the ratio of the characteristic G and 2D bands to be I(2D)/I(G) = 2.3, which confirms it to be single layer. Absence of D-peak implies graphene to be defect-free.

Lithography with resists and wet chemicals was avoided for making electrical contacts on graphene to prevent its contamination which significantly affects the Pb morphology and the graphene-Pb interface transparency. The contacts on the single layer graphene (SLG) were made by depositing Cr/Au (5/45 nm) in a van der Pauw geometry using a mechanical mask, as shown in Fig. \ref{fig:GrPb1}(b). Pb was then deposited using thermal evaporation technique. It can be observed from the SEM image in Fig. \ref{fig:GrPb1}(c) that Pb formed discrete nanometer-sized islands on graphene in the size range from 30 to 300 nm, instead of forming a uniform layer due to its poor wettability on graphene \citep{han2020disorder,han2020gate}. Thus the Pb-graphene hybrid device is a 2D random array of Josephson junctions with a distribution of island size and there is some distribution in inter-island separation as well.

While depositing Pb on graphene the substrate temperature during deposition as well as Pb deposition rate and thickness play a key role in deciding the size, size-distribution of Pb islands and also their separation \citep{yu1991coalescence,liu2015growth}. 30 nm Pb was thermally evaporated on graphene by keeping the temperature of the substrate at 71$^{\circ}$C. Pb deposited on the $\mathrm{SiO_2}$ substrate surrounding graphene, also formed distinct islands with an inter-island separation larger than that on graphene, thus preventing any electrical conduction through them. Hence the electrical conduction happens only through Pb islands that are coupled through graphene. A good quality of the interface between Pb and graphene was also ensured by depositing Pb at a high rate of 20 \AA/sec, apart from avoiding wet chemical processes.

Devices after depositing Pb were promptly mounted on a cryostat which was then cooled in a closed cycle refrigerator to its base temperature of 1.3 K. Low pass R-C filters with a 15 kHz cutoff frequency as well as high-frequency-cutoff pi-filters were installed in the measurement lines at room temperature to minimize electromagnetic noise. The transport measurement wires also go through Cu-powder filters at the base temperature to further reduce the noise interference. The four-probe transport measurements were carried out using a dc-current source as depicted in Fig. \ref{fig:GrPb1}(d). Current was varied over a limited range, from $-$20 to 20 $\mu$A, to minimize Joule heating. For the resistance measurements, the device was biased with 1 $\mu$A current of both polarities. The voltage measured from the device was amplified using a Femto voltage amplifier. Gate voltage ($V_{\rm g}$) between $-$90 to 90 V was applied to Si substrate with 10 k$ \Omega$ series resistance.

\section{Theoretical background}

\subsection{Berezinskii-Kosterlitz-Thouless model}
\label{BKT}

The BKT transition is described for ordered 2D thermodynamic systems with interaction between two vortices having a logarithmic dependence on their separation. According to the BKT model, thermal unbinding of the bound vortex anti-vortex pairs into free vortices happens above the vortex-unbinding or the BKT transition temperature $T_{\rm BKT}$. A current, small enough not to induce vortex unbinding in a 2D-SC, induces motion of free vortices due to a Lorentz force in a direction perpendicular to the current on the vortices. This motion of free vortices leads to a voltage and thus dissipation. The resistance due to the motion of free vortices in the low current limit is given by \citep{halperin1979resistive,kadin1983renormalization},
\begin{equation}
R/R_{\rm N}=a\exp\left[-2b\sqrt{(T_{\rm CO}-T)/(T-T_{\rm BKT})}\right].
\label{eq:KTequation}
\end{equation}
Here $a$ and $b$ are non-universal constants of the order of unity, $R_{\rm N}$ is the normal state resistance and $T_{\rm CO}$ is the critical transition temperature.

For a 2D-SC, the perpendicular penetration depth \cite{pearl1964current} is given by $\lambda_\perp=2\lambda^2/d$ with $\lambda$ as the penetration depth and $d$ as the thickness. Thus $\lambda_\perp$ is much larger than $\lambda$ and it can easily exceed the size of the 2D-SC samples. $\lambda_\perp$ also dictates the crossover in the separation ($r$) dependence of the inter-vortex interaction force, which varies as $1/r$, for $r < \lambda_\perp$, and as $1/r^2$ for $r > \lambda_\perp$. Thus the inter-vortex interaction energy is logarithmic for $r < \lambda_\perp$ and it decays much faster for larger $r$. Therefore, for the BKT model to be applicable to a 2D-SC and in the thermodynamic limit, the system size $L$ should be very large but at the same time it should be smaller than $\lambda_\perp$, to ensure a logarithmic inter-vortex interaction.

In a uniform array of Josephson junctions with lattice parameter $a_0$, the energy required to generate a vortex anti-vortex pair separated by $r$ is given by $2\pi E_{\rm J} \ln(r/a_{\rm 0})$ \cite{holzer2001finite}. Thus, there is finite probability for creation of such pairs at short separation and at non-zero temperatures. On the contrary, the energy required to produce a single vortex, given by $\pi E_{\rm J} \ln(L/a_{\rm 0})$, is very large for a large size array of linear dimension $L$. But, a finite size system can allow for the formation of free vortices at non-zero temperatures \cite{holzer2001finite}. Furthermore, $\lambda_\perp$ may not always be greater than the system size. For proximity-coupled Josephson junction arrays, $\lambda_\perp$ is given by \cite{newrock2000two,herbert1998effect},
\begin{equation}
\lambda_\perp(T)=\Phi_0/[2\pi\mu_0I_{\rm C}(T)].
\label{eq:lambda}
\end{equation}
Here, $I_{\rm C}$ is the critical current of each junction, which depends exponentially on temperature \cite{lobb1983theoretical} for proximity junctions. Thus, $\lambda_\perp$ can become smaller than the system size with cooling and this could happen above the anticipated $T_{\rm BKT}$ and lead to a non-logarithmic interaction between some of the vortex anti-vortex pairs that have separation more than $\lambda_\perp$. In this case, $\lambda_\perp$ will be the relevant length scale than the system size $L$ and the energy of a free vortex gets modified to $\pi E_{\rm J} \ln(\lambda_\perp/a_{\rm 0})$. Hence, in case of either of the violated condition: finite size $L$ or $\lambda_\perp < L$, it is possible for free vortices or ``finite-size vortices" to form at non-zero temperatures.

Apart from $\lambda_\perp$, another important length scale in the BKT model is the vortex correlation length $\xi_+ (T)$. For $T > T_{\rm BKT}$, $\xi_+(T) \propto \exp\{b\sqrt{(T_{\rm CO}-T_{\rm BKT})/(T-T_{\rm BKT})}\}$ \citep{halperin1979resistive,kadin1983renormalization}. $\xi_+$ represents the length scale above which vortex pairs begin to unbind, or alternatively, it is the average distance between two free vortices. As $T_{\rm BKT}$ is approached from above, $\xi_+$ grows and eventually diverges leading to no free vortices below $T_{\rm BKT}$. But for a finite size system, a cut-off is imposed on the $\xi_+$ at a temperature slightly above $T_{\rm BKT}$ and $\xi_+$ cannot grow beyond $\mathscr{L} \equiv$ min$[L,\lambda_\perp]$ \cite{schneider2014suppression}. The behavior of $\xi_+$ is also modified in the presence of a finite applied current due to current induced pair breaking effects even down to zero temperature. Thus $\xi_+$ does not diverge at $T_{\rm BKT}$ for finite currents. The finite current introduces an extrinsic length scale `$r_{\rm C}$' in the system \citep{pierson1999dynamic,newrock2000two} as well, such that the vortex pairs with separation more than $r_{\rm C}$ unbind due to applied current. In summary, unbound or free vortices may be present in a 2D-SC system at any finite temperature for several reasons including finite-size effects and finite current.

\subsection{Vortex pinning and Ambegaokar-Halperin model}
\label{AH}
The Ambegaokar-Halperin (AH) theory quantitatively models the observed small finite resistance due to thermally activated phase slippage (TAPS) in a current biased over-damped Josephson junction. For a bias-current much smaller than the critical current, the resistance of a Josephson junction at finite temperature is given by \cite{ambegaokar1969voltage},
\begin{equation}
R = R_{\rm N}[I_{\rm 0}(\gamma/2)]^{-2}.
\label{eq:Ambegaokar-Halperin}
\end{equation}
Here $R_{\rm N}$ is the normal state resistance, $I_{\rm 0}$ is the modified Bessel function of zero order and $\gamma$ is the ratio of the barrier height, in the periodic potential experienced by the phase, to $k_{\rm B}T$. The barrier height for a single junction is $E_{\rm J}=\hbar I_{\rm C}/2e$ with $I_{\rm C}$ as the junction's critical current. Moreover, the detailed current-voltage characteristic (IVC) will be non-linear as the applied current reduces the barrier for the phase-slip. The barrier vanishes as bias-current approaches the critical current $I_{\rm C}$ in a single junction.

In a uniform square 2D Josephson junction array, Rzchowski et al.\cite{rzchowski1990vortex} calculated the potential seen by a single vortex to be like an ``egg-crate", where the barriers are at the junctions between two superconducting islands and the minima lies at the junction of four superconducting islands, in case of a square array. On the application of bias current, a vortex will move in this potential, crossing the barriers at the junctions with the help of thermal energy and resulting into a phase slip of 2$\pi$ across the junction it moves through. Tinkham \cite{tinkham1988resistive} also argued that the kinetics of the driven, highly damped, thermally activated process in granular superconductors involves the same 2$\pi$-phase slip physics as the thermally activated phase motion in a single, over-damped, current-driven Josephson junction.
\begin{figure*}
	\centering
	\includegraphics[width=6.8in]{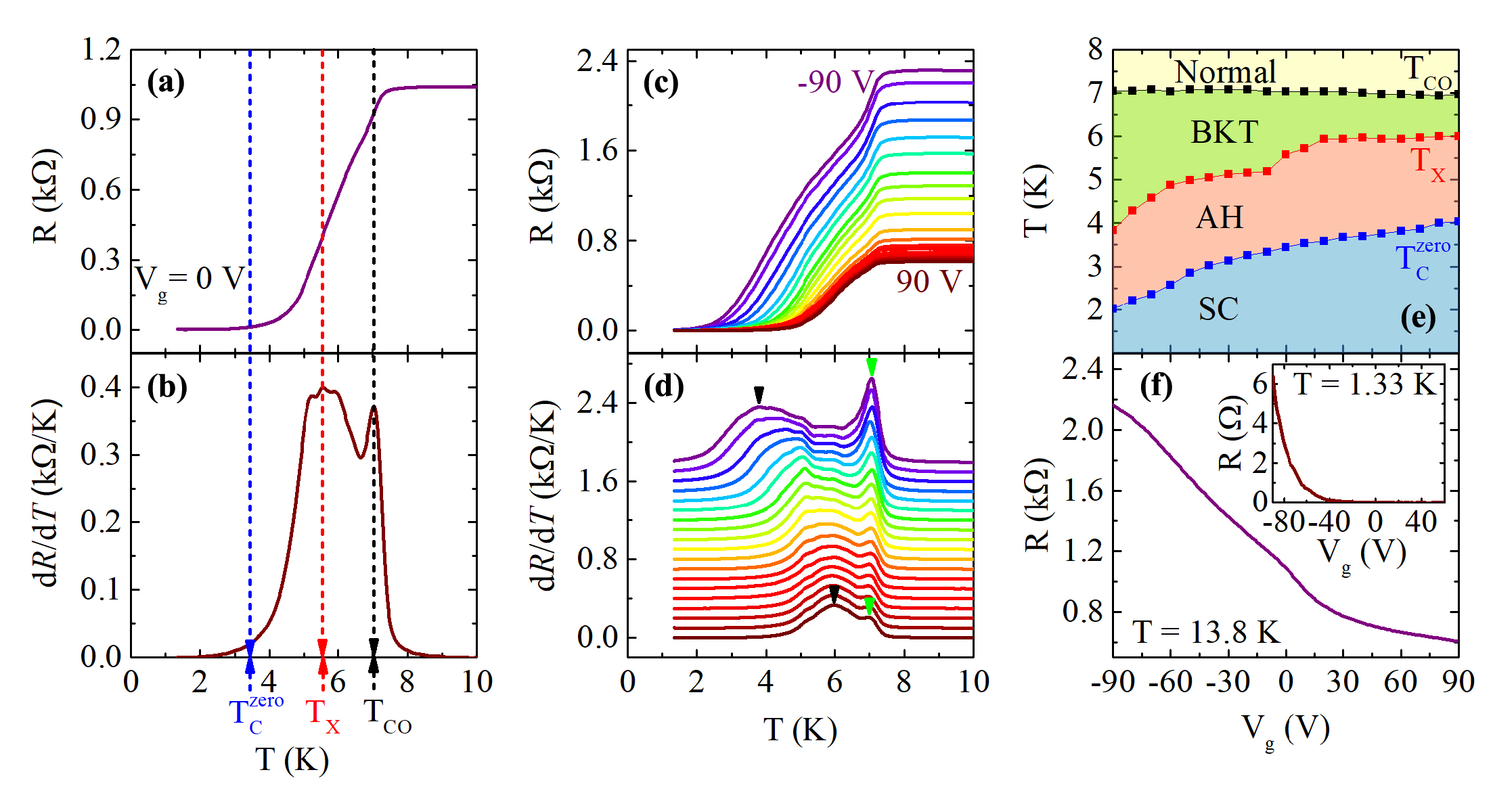}
	\caption{(a) Four probe resistance $R$ (at 1 $\mu$A bias current) and, (b) its first derivative $dR/dT$, as a function of temperature at gate voltage $V_{\rm g}$ = 0 V, illustrating the two step resistive transition. The vertical black, red and blue dashed lines mark the onset temperature $T_{\rm CO} = 7.04$ K, the crossover temperature $T_{\rm X} = 5.58$ K and the global SC temperature $T_{\rm C}^{\rm zero}= 3.44$ K, respectively. (c) and (d) show the variation in $R$ and $dR/dT$ versus $T$, respectively, with gate voltage $V_{\rm g}$. Here, $V_{\rm g}$ varies from $-$90 to 90 V (upper to lower curve) with $\Delta V_{\rm g}$ = 10 V. The curves in (d) have been uniformly shifted vertically for clarity. (e) shows the variations of $T_{\rm CO}$, $T_{\rm X}$ and $T_{\rm C}^{\rm zero}$, with $V_{\rm g}$. (f) depicts the $V_{\rm g}$ dependence of the resistance at $T=$ 13.8 K showing that the Dirac point of the hybrid system is below $-$90 V. The inset of (f) shows the variation of resistance with $V_{\rm g}$ and at 1.33 K base temperature.}
	\label{fig:GrPb2}
\end{figure*}

For a JJ array, the actual barrier seen by a vortex in a minimum is lower than $E_{\rm J}$, i.e. the single junction phase slip barrier. For a uniform square JJ array, this barrier is found to be \cite{lobb1983theoretical} close to $0.2E_{\rm J}$. The exact pre-factor of 0.2 will be different for different array geometries and for a random array there will be a distribution in barrier energies. The actual $\gamma$ would be determined by this modified barrier seen by a vortex and thus it is more appropriate to use $\gamma=0.2E_{\rm J}/k_{\rm B}T$. Using this reduced barrier it is reasonable to modify \cite{lobb1983theoretical} the finite bias current AH expression \citep{ambegaokar1969voltage,wright1991dissipation} to arrive at the array voltage $V$ as,
\begin{align}
V= 2&(0.2I_{\rm C}R_{\rm N}) \sqrt{1-x^2}\nonumber\\
\times&\exp{[-\gamma (\sqrt{1-x^2} + x\sin^{-1}x)]} \sinh{(\pi \gamma x/2)}.
\label{eq:Ambegaokar-Halperin_AnalyticExp}
\end{align}
Here, $x = 5I/I_{\rm C} < 1$ and $\gamma=0.2E_{\rm J}/k_{\rm B}T>1$. The above Eq. \ref{eq:Ambegaokar-Halperin} is consistent with Eq. \ref{eq:Ambegaokar-Halperin_AnalyticExp} in $x=0$ limit. It is noteworthy that the low bias transport characteristics of a JJ array are not dictated by the sum of the critical currents of all the junctions in the array's width. It is rather dictated by a critical current which is even smaller than that of a single junction of the array. The reduction in critical current due to de-phasing induced by trapped vortices is well known in Josephson devices.

In a JJ array, a finite density of thermally generated vortices and anti-vortices can exist at non-zero temperature and in zero magnetic field. These can escape at certain rate, under an applied bias current, from the edges of a finite size JJ array and then get replaced by newly generated pairs in the bulk. The pairs' escaping at a fixed rate amounts to a steady voltage. The barrier for the escape of a vortex from an edge can be expected to be about $0.2E_{\rm J}$ while that for the entry of a vortex will be close to $E_{\rm J}$. This `vortex and anti-vortex escape' mode of flux transport is far more probable under thermal fluctuations than the entry of a vortex from an edge followed by its traversal through the array and then escape from the opposite edge. The latter mode dominates in single or few junction devices like SQUIDs.

Tinkham \cite{tinkham1988resistive} defined a suitable activation barrier replacing $E_{\rm J}$ and used a generalized form of $\gamma$ as $\gamma=A(1-T/T_{\rm CO})^{3/2}$ with $A$ as an empirical constant. Later, Bhalla et al. \citep{bhalla2004dissipation,bhalla2007vortex} used this generalized form in explaining dissipation in the polycrystalline $\mathrm{YBa_2Cu_3O_{7-\delta}}$ sample. Because of the structural irregularity in their system owing to the varied orientations and sizes of Josephson junctions, they used a modified expression: $\gamma=A(1-T/T_{\rm CO})^m$, replacing the exponent 3/2 by a variable exponent `$m$'. The constant $A$ included an unspecified dependence of $\gamma$ on magnetic field $B$. Accordingly, the zero-bias resistance Eq. (\ref{eq:Ambegaokar-Halperin}) gets modified to:
\begin{equation}
R = R_{\rm N}\{I_{\rm 0}[A(1-T/T_{\rm CO})^m/2]\}^{-2}.
\label{eq:Ambegaokar-Halperin_1}
\end{equation}
For our graphene-Pb hybrid devices both $A$ and $m$ can be expected to depend on the gate voltage $V_{\rm g}$

\section{results and analysis}

\subsection{Gate dependent resistance measurements}
\label{RT_Vg}

Fig. \ref{fig:GrPb2}(a) displays the measured resistance with temperature at $V_{\rm g}$ = 0 V for the graphene-Pb hybrid sample. It shows a broad resistive transition with cooling as the resistance $R$ drops gradually starting from a superconductivity onset temperature of about 7 K, which is close to the SC critical temperature of the bulk lead. The broadened R(T) curve displaying a two-step transition to zero resistive state, is characteristic of granularity in SC \citep{eley2012approaching,zhang2014global,derevyanko2017phase}.  The first derivative of resistance with respect to temperature, i.e. $dR/dT$, see Fig. \ref{fig:GrPb2}(b), exhibits two peaks. The higher temperature narrow peak corresponds to the onset temperature $T_{\rm CO}$ which is associated with the superconductivity inside the Pb islands. Below $T_{\rm CO}$, a weak Josephson coupling with energy $E_{\rm J}$ sets-in leading to a weak phase coherence between islands with phase-fluctuations proliferating as vortices.

With cooling, $E_{\rm J}$ increases and a relatively less steep reduction in resistance, marked by a broader peak in $dR/dT$, as compared to that at $T_{\rm CO}$, is observed. The temperature corresponding to this peak marks a crossover between two regimes, as discussed later, and thus it is denoted as $T_{\rm X}$. This peak also has some fine structure, which presumably arises from a distribution in $E_{\rm J}$ values and is influenced by the detailed island distribution in this finite size sample. As the temperature is further lowered, global phase coherence is achieved leading to a macroscopic SC state. The temperature $T_{\rm C}^{\rm zero}$ at which macroscopic SC state is achieved, is defined at 1\% of $R_{\rm 10 K}$.

\begin{figure}[h!]
	\centering
	\includegraphics[width=3.4in]{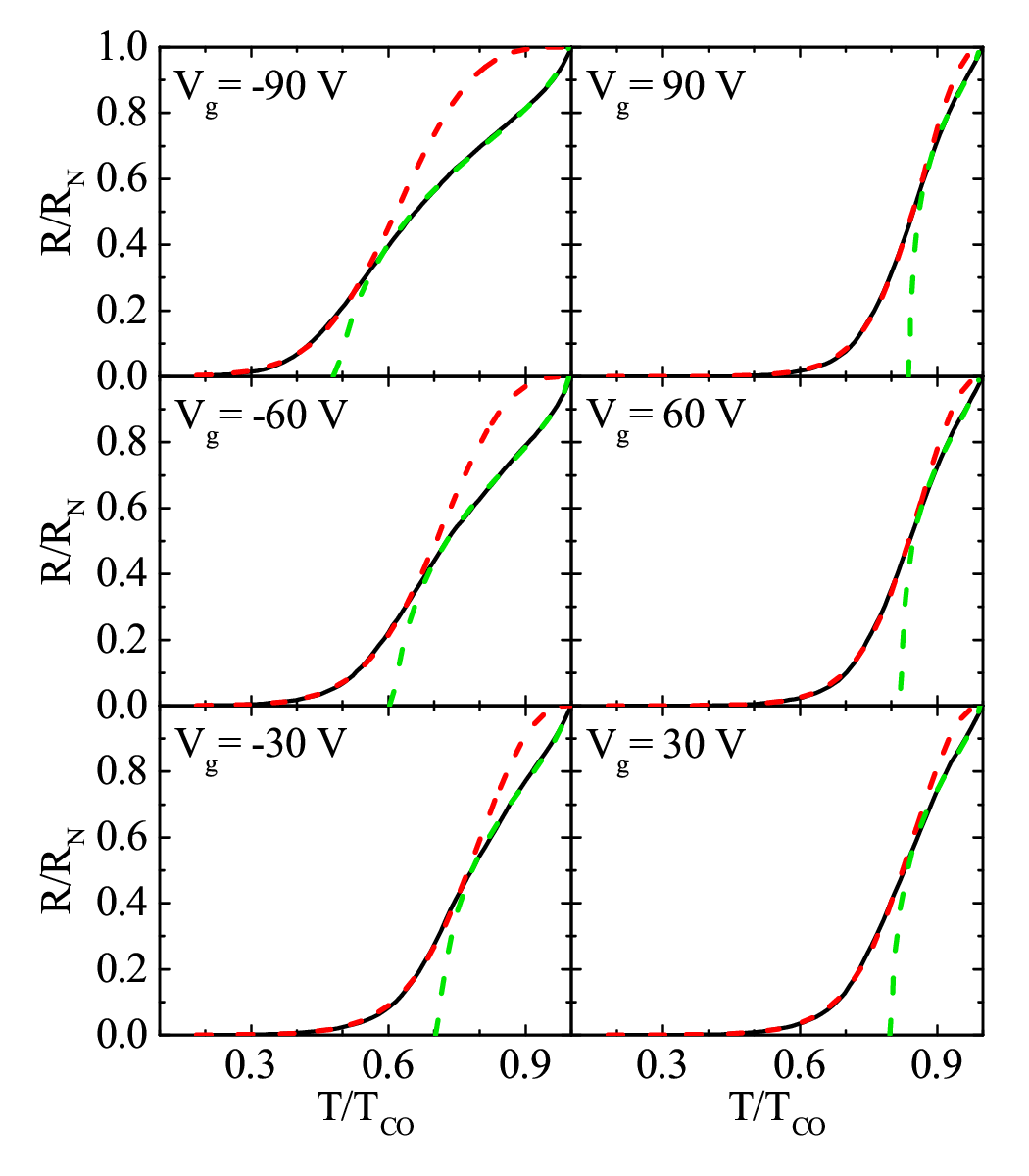}
	\caption{Normalized resistance as a function of normalized temperature at $V_{\rm g} =$ $-$90, $-$60, $-$30, 30, 60 and 90 V. The green and red dashed lines are the fits using BKT Eq. (\ref{eq:KTequation}) and AH Eq. (\ref{eq:Ambegaokar-Halperin_1}), respectively.}
	\label{fig:GrPb3}
\end{figure}

There is significant variation in $R(T)$ and $dR/dT$ curves with $V_{\rm g}$ as shown in Fig. \ref{fig:GrPb2}(c,d). The carrier density in this graphene, which is electron-doped due to the interface traps and presence of Pb, is the lowest at $V_{\rm g}=-90$ V and it increases monotonically with $V_{\rm g}$. The Dirac point in this sample is not accessible due to significant electron doping as seen from the normal state resistance variation with $V_{\rm g}$ in Fig. \ref{fig:GrPb2}(f). The Josephson coupling between Pb islands, mediated by graphene, is expected to increase with $V_{\rm g}$ due to increase in carrier density in graphene. As a result, when $V_{\rm g}$ is increased from -90 to +90 V, the $T_{\rm C}^{\rm zero}$ increases from about 2 K to 4 K although $T_{\rm CO}$ remains nearly the same as shown in Fig. \ref{fig:GrPb2}(e). Thus the transition region, between $T_{\rm CO}$ and $T_{\rm C}^{\rm zero}$, widens with reducing $V_{\rm g}$ as $T_{\rm X}$ reduces. Interestingly, with reducing $V_{\rm g}$ or $E_{\rm J}$, the peak in $dR/dT$ at $T_{\rm CO}$ becomes more pronounced indicating a much sharper onset of SC. At the same time the peak at $T_{\rm X}$ becomes broader with reducing $V_{\rm g}$ and the fine-structure in it becomes clearer.

Fig. \ref{fig:GrPb3} shows the plots of $R/R_{\rm N}$ versus $T/T_{\rm CO}$ at several $V_{\rm g}$ values. The green dashed lines are the least-square fits up to $T_{\rm X}$ using Eq. (\ref{eq:KTequation}), yielding $a$, $b$ and $T_{\rm BKT}$ as the fitting parameters. $R_{\rm N}$ is taken to be 90\% of $R_{\rm 10 K}$ value at the corresponding $V_{\rm g}$. Since the BKT fit works well for $T_{\rm CO}>T>T_{\rm X}$, the vortex dynamics in this regime is dominated by logarithmic inter-vortex interactions and, presumably, $\lambda_\perp$ exceeds the sample size. As the temperature reduces, $\lambda_\perp$ will decrease as $I_{\rm C}$ increases, see Eq. \ref{eq:lambda}, and at some temperature $\lambda_\perp$ may reduce below the sample size. Then the inter-vortex interaction energy for vortex pairs with separation farther than $\lambda_\perp$ will reduce significantly and their interaction will not play much role. Deviation from Eq. (\ref{eq:KTequation}) is observed below $T_{\rm X}$ which is slightly more than the $T_{\rm BKT}$, as seen in Fig. \ref{fig:GrPb3}. The system should go to zero resistance state below $T_{\rm BKT}$ but it starts deviating from BKT fit above $T_{\rm BKT}$ and with a finite resistive tail. This resistive tail implies dissipation in the system due to some sort of free vortices which are present, presumably, due to finite-size effects and current induced unbinding as discussed in section \ref{BKT}.

\begin{figure}[h!]
	\centering
	\includegraphics[width=3.4in]{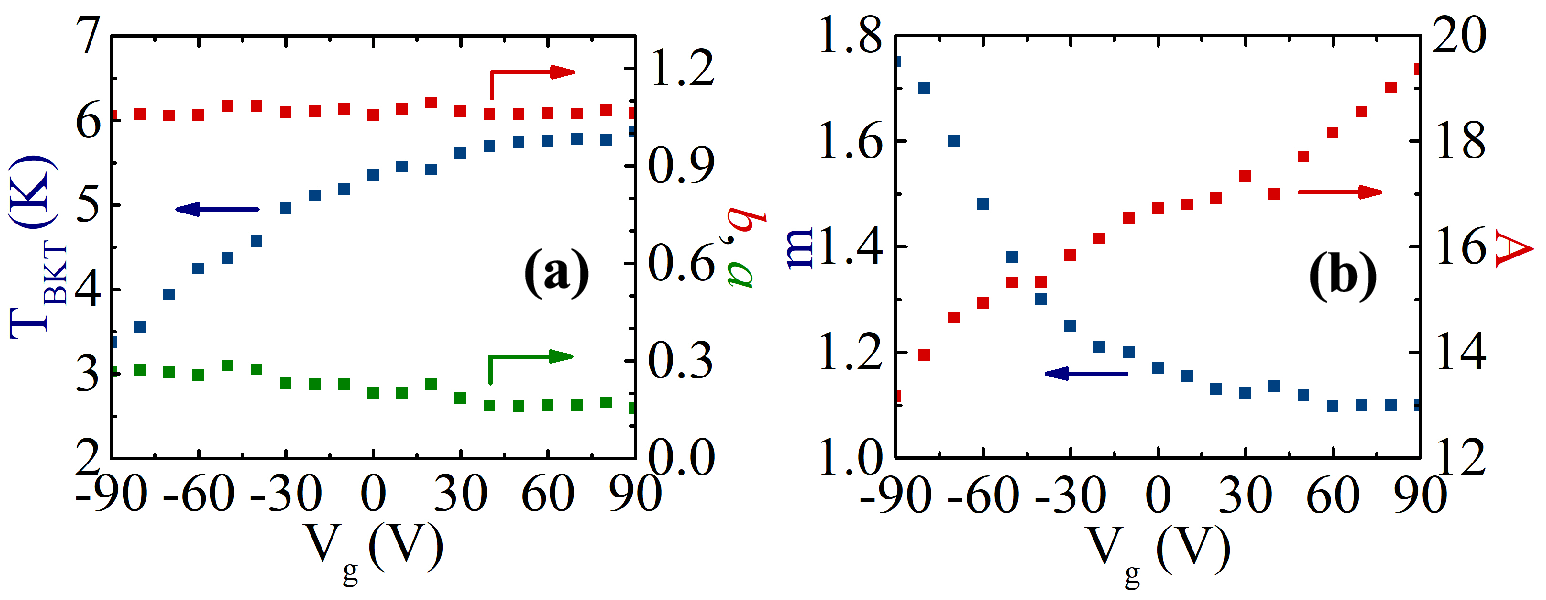}
	\caption{Fitting parameters (a) $a, b$ and $T_{\rm BKT}$ for $T>T_{\rm X}$, and (b) $m$ and $A$ for $T<T_{\rm X}$, plotted as a function of $V_{\rm g}$. }
	\label{fig:GrPb4}
\end{figure}

As the system cools down, $E_{\rm J}$ increases and more Pb islands become phase coherent due to proximity induced superconductivity in graphene \cite{feigel2008proximity}. When $E_{\rm J}$ exceeds $k_{\rm B}T$ substantially a pinning dominated regime may occur. The pinning here does not refer to the intra-island pinning, which will cost significantly more energy. The region between the Pb islands on graphene forms a complex network of channels for vortices to move along and as discussed earlier the vortices will have lower energy, or a tendency to get pinned, at the border of many islands as compared to the border of two islands. The detailed energy landscape for a vortex in the JJ arrays is dictated by $E_{\rm J}$ and the distribution of Pb islands. Eq. (\ref{eq:Ambegaokar-Halperin_1}) is fitted to the resistive tails and below $T_{\rm X}$. As discussed earlier, the unspecified $V_{\rm g}$ dependence of $\gamma$ is included in $A$ and $m$. With further cooling, the vortices fail to get activated from pinning sites and thus below $T_{\rm C}^{\rm zero}$, these vortices are localized with a static phase difference at low bias between the Pb islands resulting into the zero resistance state.
\begin{figure}[h!]
	\centering
	\includegraphics[width=3.4in]{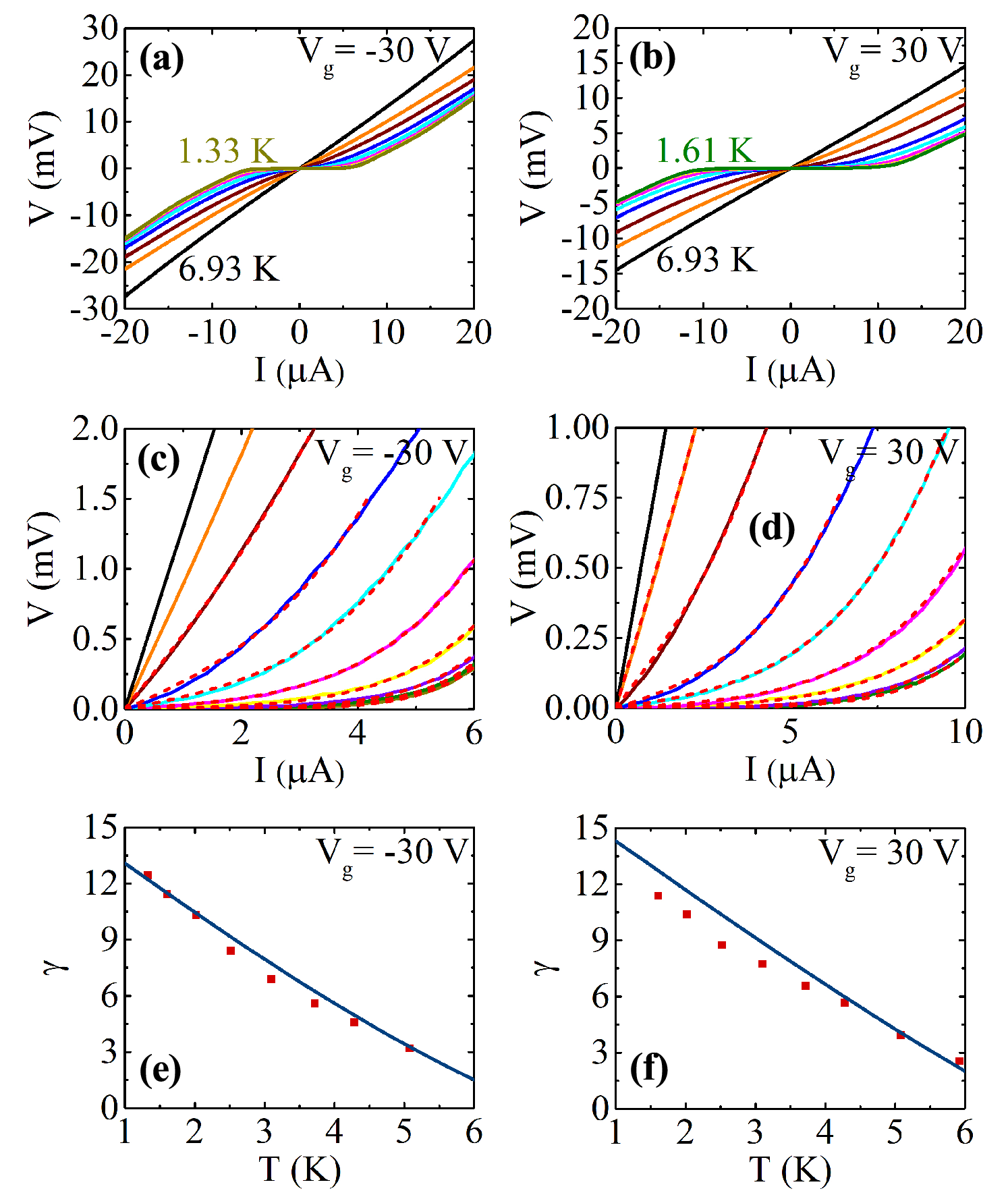}
	\caption{(a) V-I characteristics at temperatures $T$ = (1.33, 1.61, 2.02, 2.52, 3.10, 3.72, 4.28, 5.08 K) $< T_{\rm X} $ and (5.92, 6.93 K) $> T_{\rm X} $ for $V_{\rm g}$ = $-$30 V. (b) V-I characteristics at temperatures $T$ = (1.61, 2.02, 2.52, 3.10, 3.72, 4.28, 5.08, 5.92 K) $< T_{\rm X} $ and 6.93 K $> T_{\rm X} $ for $V_{\rm g}$ = 30 V. These curves are zoomed in (c) and (d). Red dashed lines are the fits to Eq. (\ref{eq:Ambegaokar-Halperin_AnalyticExp}) for V(I) curves at $T < T_{\rm X}$. (e) and (f) show temperature dependence of $\gamma$ obtained by fitting V(I) (red symbols) and that deduced from $A$ and $m$ (blue lines) used in fitting $R(T)$ curves below $T_{\rm X}$.}
	\label{fig:GrPb5}
\end{figure}

Eq. (\ref{eq:Ambegaokar-Halperin_1}) breaks down for $T > T_{\rm X}$, as the inter-vortex interaction starts dominating over the pinning, and leads to the BKT regime. For $T< T_{\rm X}$ pinning dominates over inter-vortex interaction. Fitting parameters from the two models are plotted in Fig. \ref{fig:GrPb4}, as a function of $V_{\rm g}$. It can be observed, in Fig. \ref{fig:GrPb4}(a), that the parameters $a$ and $b$ of Eq. (\ref{eq:KTequation}) are of the unity order as stated in the theory \cite{halperin1979resistive}. The monotonic increment of $T_{\rm BKT}$ with increase in $V_{\rm g}$ is also consistent with literature \citep{feigel2008proximity,han2014collapse}. In Fig. \ref{fig:GrPb4}(b), the fitting parameter $A$ of Eq. (\ref{eq:Ambegaokar-Halperin_1}) increases with $V_{\rm g}$. This is expected as Josephson coupling grows with $V_{\rm g}$. However, the exponent $m$ decreases slightly with increase in $V_{\rm g}$ and then saturates. This could arise from the increase in carrier density inhomogeneity in graphene and thus in $E_{\rm J}$ as one approaches the Dirac point. The island size and their separation is the other source of inhomogeneity that may not change with much $V_{\rm g}$.

\subsection{V-I Characteristics}
\label{IV}
Figure \ref{fig:GrPb5}(a,b) show the measured voltage as a function of the bias current in four-probe configuration at different temperatures for $V_{\rm g}=-30$ and 30 V, respectively. The non-hysteretic, smooth and without any abrupt jump in voltage as a function of current suggest that the system is in over-damped limit and with negligible heating effects. Further, there is a non-zero slope at zero bias current, particularly, for temperatures close to $T_{\rm X}$, as shown in the enlarged Figs. \ref{fig:GrPb5} (c,d). The red dashed curves in these figures are the fits obtained using the Eq. \ref{eq:Ambegaokar-Halperin_AnalyticExp} with the fitting parameter $\gamma$ as shown in Fig. \ref{fig:GrPb5} (e,f) as red symbols as a function of temperature. The solid lines show $\gamma=A(V_{\rm g})(1-T/T_{\rm CO})^m$ with the $A$ and $m$ taken from Fig. \ref{fig:GrPb4}(b) at respective $V_{\rm g}$ values that were obtained by fitting Eq. (\ref{eq:Ambegaokar-Halperin_1}) to the experimental R(T) curves. The agreement between the two $\gamma$ provides significant support on the applicability of the AH model. Further, the decrease of $\gamma$ with increasing temperature is consistent with reduction in Josephson coupling with temperature leading to easier flux transport.
\begin{figure}[h!]
	\centering
	\includegraphics[width=3.4in]{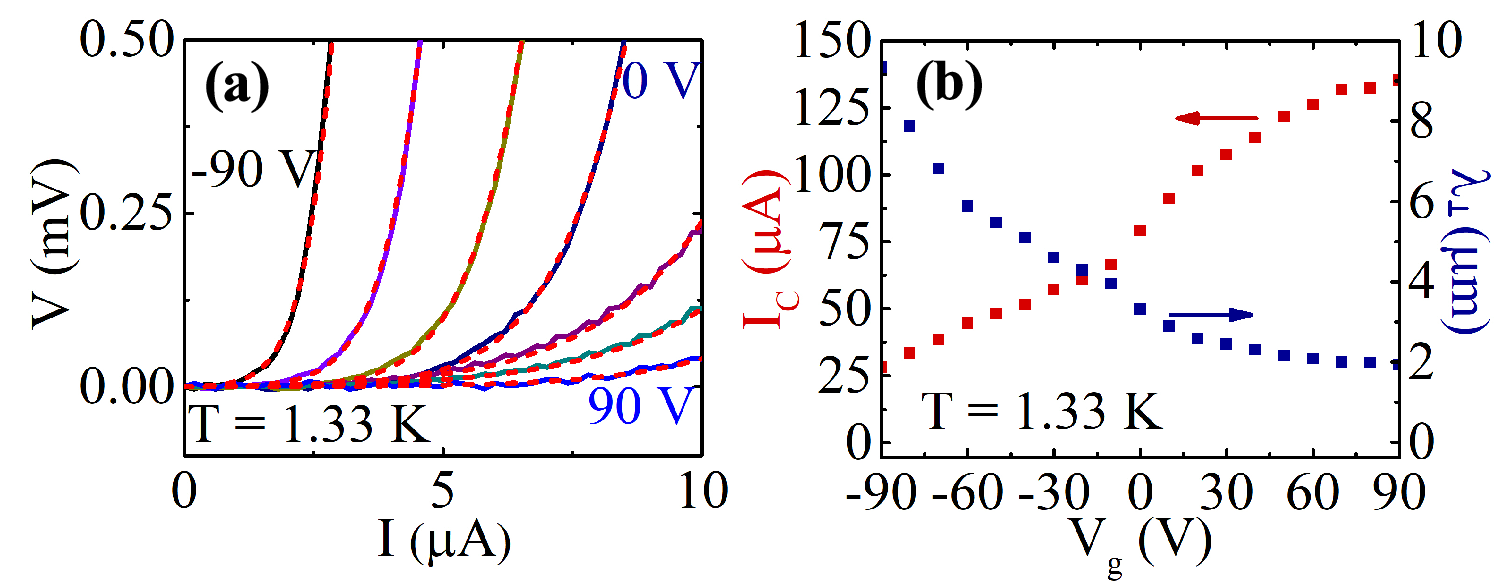}
	\caption{(a) V-I characteristics for $V_{\rm g}= -90, -60, -30$, 0, 30, 60 and 90 V at $T$ = 1.33 K. The red dashed lines are the fit to Eq. (\ref{eq:Ambegaokar-Halperin_AnalyticExp}). (b) The red symbols show the $V_{\rm g}$ dependence of $I_{\rm C}$ obtained from the fits in (a) and the blue symbols show the $V_{\rm g}$ dependence of screening length deduced using the $I_{\rm C}$ and Eq. \ref{eq:lambda}.}
	\label{fig:GrPb6}
\end{figure}

Figure \ref{fig:GrPb6}(a) shows the measured V(I) curves for a wider $V_{\rm g}$ range and at $T$ = 1.33 K. The fits with Eq. (\ref{eq:Ambegaokar-Halperin_AnalyticExp}) are made to identify the other fitting parameter $I_{\rm C}$, which is plotted as red dots in Fig. \ref{fig:GrPb6}(b) as a function of $V_{\rm g}$. Note that the actual barrier for vortex transport is only 0.2$E_{\rm J}$ or the effective mean $I_{\rm C}$ of a junction of the array is five times higher than that reflected in the $V(I)$ curves. Eq. \ref{eq:lambda} is used to deduce the value of $\lambda_{\perp}$ from these $I_{\rm C}$ values as a function of $V_{\rm g}$ at 1.33 K and as shown in Fig. \ref{fig:GrPb6}(b) as blue symbols. These values are smaller than sample's linear dimensions and these are expected to increase substantially with increasing temperature. This justifies the inapplicability of BKT physics at low temperatures as discussed earlier and the applicability of the AH model to capture the dynamics of vortices with interaction-range shorter than sample size and with more pinning.

\section{discussion and conclusion}
The above results and analysis show that the energy $E_{\rm J}$ plays two roles in JJ arrays: 1) it dictates the barrier for vortex motion and, 2) it controls the screening length $\lambda_{\perp}$. A higher $E_{\rm J}$ leads to larger phase stiffness and increases the height of the barriers faced by vortices but it leads to a shorter screening length $\lambda_{\perp}$ which dictates the vortex size and the range of inter-vortex interaction. When $\lambda_{\perp}$ reduces below the sample size, with reducing temperature, the inter-vortex interaction weakens and the vortex dynamics is than dominated by increased pinning due to the larger barriers. The BKT model below $T_{\rm X}$ would anticipate a low voltage as the free vortex density reduces; however, the $\lambda_{\perp}$ becoming smaller than sample size leads to much less inter-vortex interaction between far away vortices and so there can be more free vortices than BKT expectation. This explains higher voltages below $T_{\rm X}$ than that expected from BKT. An interesting coincidence is that the $T_{\rm X}$ is higher than the expected $T_{\rm BKT}$ for all $V_{\rm g}$ values. As a result, the free vortex density does not vanish before the AH dominated regime takes over and thus a monotonic reduction in resistance is observed with cooling. For $T>T_{\rm X}$, the AH model becomes inapplicable as the the barrier faced by vortices becomes smaller making vortices more mobile and at the same time the actual hindrance comes from the inter-vortex interaction as the $\lambda_{\perp}$ increases making more vortices interact. The AH model anticipates more voltage or resistance above $T>T_{\rm X}$ with reduction in vortex hindrance but than the inter-vortex interaction impedes the vortex motion reducing the voltage below the AH expectation, as observed.

It is also noteworthy that the observed crossover between the two regimes at $T_{\rm X}$ shows a smooth matching at this temperature and without any noticeable regime that does not fit either of these two models. Further, this is the case for all studied $V_{\rm g}$ values. This makes us believe that a single model should be able to describe both the regimes. In fact, Lobb et. al. \cite{lobb1983theoretical} proposed a model for JJ arrays that effectively combined the AH model and the BKT model with the former dictating the bias current induced drift rate of the free vortices and the later dictating the density of free vortices. Eventually, they argued that the drift rate does not change so drastically with temperature as compared to the free vortex density and thus in the temperature range of interest, the BKT model was assumed to be valid. In our studied sample, the AH model fits the resistance below $T_{\rm X}$ implying that the change in free vortex density with temperature does not dominate below $T_{\rm X}$.

In conclusion, the Pb-graphene hybrid system exhibits gate-tunable two-step transition to the zero resistance state as it is cooled below $T_{\rm CO}$, i.e. when the islands become superconducting. The initial dissipation is from the motion of the free vortices and it fits well with the BKT model. But with cooling, the screening length $\lambda_\perp$ decreases and $E_{\rm J}$ increases and thus the pinning dominates below a crossover temparature $T_{\rm X}$. This regime fits well with the AH theory. The increase in $T_{\rm X}$ with $V_{\rm g}$ provides a tunability in this crossover. The overall transition width increases with reducing $V_{\rm g}$ as $T_{\rm CO}$ remains the same while $T_{\rm X}$, $T_{\rm BKT}$ and $T_{\rm C}^{\rm zero}$ decrease.
Lastly, the experiment has enabled us to acquire a deeper comprehension of the dissipation in the two domains of the resistive transition.

\section*{ACKNOWLEDGMENTS}
We express our gratitude to Satyajit Banerjee and Pratap Raychaudhuri for discussions. We acknowledge SERB-DST of the Government of India and IIT Kanpur for financial support.

\end{document}